# AFM Cantilever Magnetometry for Measuring Femto-Nm Torques Generated by Single Magnetic Particles for Cell Actuation

Maria V. Efremova[1*], Lotte Boer[1*], Laurenz Edelmann[1], Lieke Ruijs[1], Jianing Li[1], Marc A. Verschuuren[2], and Reinoud Lavrijsen[1]

* equal contribution

[1]Department of Applied Physics, Eindhoven University of Technology, P.O. Box 513, 5600 MB Eindhoven, The Netherlands

[2]SCIL Nanoimprint Solutions, High Tech Campus 11, 5656EA Eindhoven, the Netherlands

**Abstract.** Particles with high anisotropy in their magnetic properties and shape are of increasing interest for mechanobiology, where transducing a remotely applied magnetic field vector to a local mechanical response is crucial. An outstanding challenge is quantifying the mechanical torque of a single nanoparticle, typically in the range of atto- to femto-Newton-meters (Nm). The magneto-mechanical torque manifests due to a misalignment of the external magnetic field vector with the built-in magnetic anisotropy axis, as opposed to a magnetic force, and complicates the measurement scheme. In this work, we developed a method using a commercially available Atomic Force Microscopy setup and cantilevers to quantify the torque generated by a single synthetic antiferromagnetic (SAF) nanoplatelet with high perpendicular magnetic anisotropy. Specifically, we measured $1.6 \pm 0.6 \cdot 10^{-15}$ Nm torque while applying 373±5 mT field at 12±2° degrees to the built-in anisotropy axis exerted by a single circular SAF nanoplatelet with 1.88 μm diameter and 72 nm thickness, naively translating to a ≈ 1.7 nN maximum force at the nanoplatelet apex. This measured torque and derived force of the SAF nanoplatelets is strong enough for most applications in mechanobiology; for example, it can be used to rupture (cancer) cell membranes. Moreover, SAF nanoplatelets open a route for easy tuning of the built-in magnetic anisotropy and size, reducing the torque and allowing for small mechanical stimuli for ion channel activation. This work presents a straightforward and widely applicable method for characterizing magnetic particles' mechanical transduction, which is applied to SAF nanoplatelets with a high PMA.

Spatiotemporal application and modulation of mechanical forces in biological systems are of great interest for future therapies, including cancer immunoengineering.[1–4] Since a pioneer study by Kim *et al.* in 2010,[5] disc-shaped structures with high magnetic and shape anisotropy have been widely investigated as micro/nanotransducers of a remotely applied magnetic field into a mechanical response.[6–8] This effect is achieved by a physical rotation of magnetic particles anchored at cell membranes or other organelles under e. g. an oscillating magnetic field, and can be utilized for mechanical destruction of cancer cells, triggering their programmable death.[9,10]

Locally, cancer cells interact with particles one-on-one.[11] Therefore, the potential of a specific magnetically anisotropic structure for magneto-mechanical actuation must be evaluated on the level of single particles.

Cell membrane rupture typically requires forces between 100 and 1000 pN.[12] In a situation when these forces are achieved via a mechanical rotation of magnetic nano/microdiscs to align their magnetic easy axis[13–15] or easy plane[5,16,17] with an external magnetic vector, concomitant mechanical torque $\vec{T}$ is generated.[18] Its value can be found as $\vec{T} = \vec{r} \times \vec{F}$, where is $\vec{F}$ is an applied force, and $\vec{r}$ is its arm, corresponding to the radius of the magnetic particle. If we assume the 1000 pN ($10^{-9}$ N) force to be generated by a μm-radius ($10^{-6}$ m) disc, this will lead to a maximum of ≈ $10^{-15}$ Nm (fNm) torque.

To quantify such a small torque generated by a single particle (as opposed to a magnetic force due to magnetic field gradients) is not straightforward when using a commercially available setup to make the measurement technique accessible. For example, the particle can be attached to some form of cantilever. When a magnetic field is applied, the torque generated by the interaction of the particle with the magnetic field is exerted on the cantilever, resulting in bending and a change in the mechanical resonance frequency. The cantilever deflection can then be read out via different methods using capacitances,[19–21] piezoelectric cantilevers,[22–24] and optical setups.[25,26] However, most of these methods require calibration with known torques to determine the absolute torque value, except for the method described in Adhikari *et al.*, 2012.[25] There, a bulk iron sample is attached to a rectangular cantilever and placed in a magnetic field. The laser beam is reflected from the cantilever and impinges onto a position-sensitive detector. The measured value is the cantilever deflection, which is directly related to the mechanical torque via the cantilever bending.[25,27–29] A simpler version of this technique that does not require a dedicated setup can be realized using a commercially available atomic force microscope (AFM), which is designed to measure forces down to the pN scale[30] and offers a variety of available cantilevers.

In this paper, we present a method using an AFM setup and commercially available cantilevers to quantify the torque generated by a single synthetic antiferromagnetic nanoplatelet (SAF NP) with high perpendicular magnetic anisotropy (PMA).[31,32] Due to the PMA, it is energetically favorable for the magnetic easy axis to be aligned with the symmetry axis of the platelet, which is perpendicular to the platelet itself (hard plane). Therefore, in a sufficiently large rotating or alternating magnetic field, the out-of-plane SAF NPs physically rotate, when free, until their easy axis is collinear with the direction of the external magnetic field vector. Compared to vortex structures and in-plane SAF NP with an easy plane and hard axis, out-of-plane SAF NP with strong PMA have a uniaxial anisotropy (easy axis, hard plane) making them efficient torque transducers.[15]

The basic working principle of AFM is shown in **Fig. 1(a)** in black and white, where the AFM signal on the photodetector stems from the laser reflection from the back side of the cantilever scanning a surface.

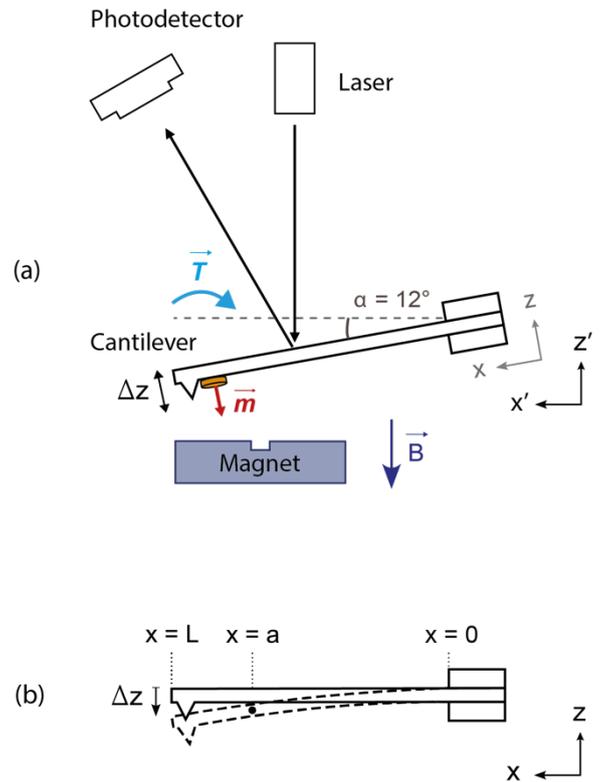

FIG. 1. The principle of the AFM cantilever magnetometry. (a) Schematic representation of the standard AFM setup (in black and white) with the modifications implemented in this study (in color), not to scale. A single SAF NP (yellow) is deposited on the tip side with its magnetic moment $\vec{m}$ oriented out-of-plane (red), a GMW 5203 vertical projected field magnet generates a magnetic field $\vec{B}$ (dark blue), and the torque $\vec{T}$ (light blue) is induced in SAF NP if $\vec{m}$ and $\vec{B}$ are not parallel, which leads to the cantilever deflection Δz. A coordinate system x'-z' is associated with the horizontal plane of the AFM setup and the measured deflection direction, while a coordinate system x-z is associated with the cantilever itself that makes an angle α = 12±2° with the horizontal plane; (b) schematic depiction of the bending of a cantilever with the length *L* due to point torque of the SAF NP applied at $x = a$. Δz is the cantilever deflection in the coordinate system x-z.

To utilize the AFM setup for measuring torques from a single SAF NP, we use a Bruker Dimension Edge AFM as shown in **Fig. 1(a)** (more details are in the **supplementary material**). The SAF NP is placed on an AFM cantilever, which is mounted on the AFM scan head via a non-magnetic cantilever holder under an angle α = 12±2° with the horizontal plane. To exclude the force that the NP would generate in a gradient magnetic field and measure a pure torque, a homogeneous magnetic field at the position of the NP is required. This is implemented via a

GMW 5203 vertical projection field magnet underneath the cantilever. Due to the shape of the magnetic pole, as well as the flux profiles inside and outside of the dent in the magnet, there is a point above the pole (3.5 mm) where the field is uniform.[33] When the vector of the field of an electromagnet $\vec{B}$ and the magnetization vector $\vec{m}$ of the SAF NP are not parallel, a torque $\vec{T}$ in the SAF NP is induced, which, in turn, acts on the cantilever.

An MLCT-B cantilever from Bruker (see the **supplementary material** for more details) was chosen for the experiments due to its rectangular shape and low spring constant. The former allows for easy calculation of the deflection for a given torque, while the latter results in a pronounced deflection, which is beneficial for the setup sensitivity. We calibrated the spring constant of MLCT-B cantilevers using the Sader method (see **supplementary material** for more details).

The measured deflection signal can be directly related to the mechanical torque through the cantilever bending. We consider the bending of a cantilever due to a force or torque (see the **supplementary material** for more details) and the coordinate system related to the AFM cantilever where $x = 0$ is the clamped end and $z = 0$ is the cantilever top surface in an undisturbed state (**Fig. 1(b)**). We take the torque of the SAF NP located at $x = a$ as a point torque applied at $x = a$, which bends the cantilever. This is a valid approximation as the radius of the NP is $\approx 100$ times smaller than the cantilever length. Using the relation between the bending of the cantilever and the applied torque,[25] as well as the equation for the point torque (see **supplementary material** for details), we find:

$$z(x) = z_a + \frac{dz}{dx}|_{x=a} (x - a) = \frac{T}{2EI}(2x - a^2) = \frac{3T}{2kx_{tip}^3}(2x - a^2) \quad (1),$$

With $T$ as the applied torque, $E$ – Young's modulus of the cantilever, $I$ – its geometrical moment of inertia, $x_{tip}$ as the position of the tip on the cantilever and $k$ as the cantilever spring constant. From equation (1), we find the maximum deflection for single SAF NPs to be on the nm scale, hence easily detected by a typical AFM (see **supplementary material** for more details).

From the measurement principle in **Fig. 1**, it is clear that the sample should have a preferential orientation of the magnetization to generate the torque within a given magnetic field geometry. To this end, we chose to work with out-of-plane SAF NPs with strong PMA achieved via the basic stack unit containing two magnetic layers of $Co_{60}Fe_{20}B_{20}$ repeated five times to increase the total magnetization[31] having the composition (layer thicknesses in nm) Ta(4)/Pt(2)/$Co_{60}Fe_{20}B_{20}$(0.8)/Pt(0.4)/Ru(0.8)/Pt(0.4)/$Co_{60}Fe_{20}B_{20}$(0.8)/Pt(2) presented in **Fig. 2(a)**.

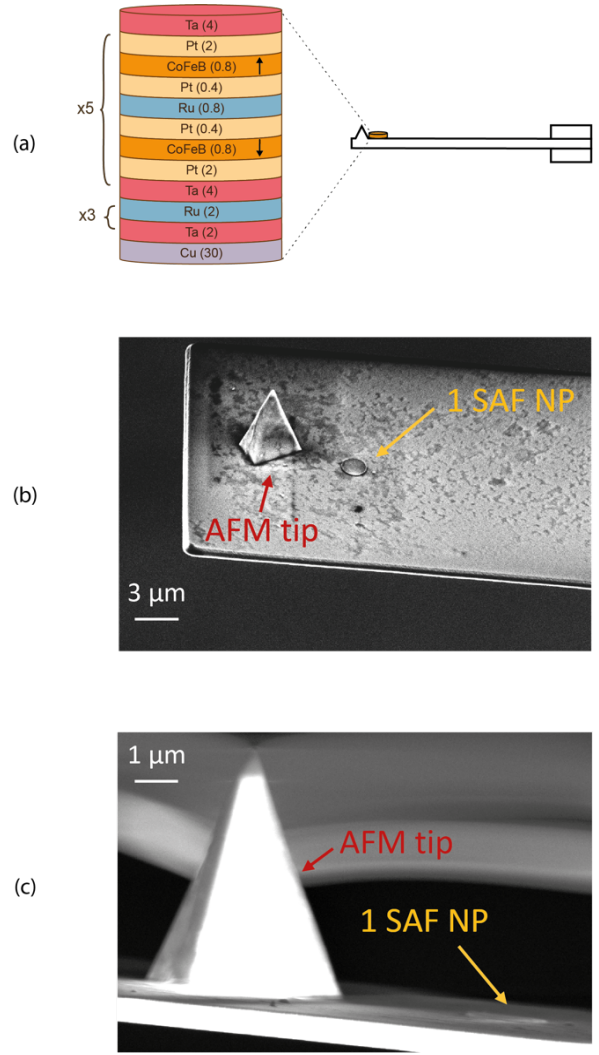

**FIG. 2.** Synthetic antiferromagnetic nanoplatelets (SAF NPs) investigated by torque magnetometry in the current study. (a) Schematic representation of the SAF NP deposited on the AFM cantilever; the inset shows a complete metallic layer stack of the SAF NP (thickness of the layers in nm); (b), (c) scanning electron microscopy

images of a single SAF NP deposited on an AFM cantilever MLCT-B in the vicinity of the tip: top view (b), side view (c).

Starting with out-of-plane SAF thin films and as described in Li et al., 2022,[31] we fabricated 1.88 μm-diameter NPs with 72 nm thickness. For the measurements, a single SAF NP was put on the tip side of Bruker MLCT-B cantilever (**Figs. 2(b)** and **2(c)**) close to the tip itself by dropcasting a water suspension of SAF NPs (see **supplementary material** for details). **Fig. 2(c)** confirms that there is only one SAF NP (cf. the orange arrow) on the cantilever, and not a stack of NPs.

To measure the magneto-mechanical torque, we explore the scheme with a pulsed field (**Fig. (3)**). Two 250 ms pulses with the same magnetic field strength are applied, in between which the magnetic field is turned off. We register the deflection of the AFM cantilever with a single SAF NP vs. the deflection of the nominally identical empty cantilever, and both datasets are averaged over 300 measurements (see the **supplementary material** for more details). Their difference is plotted in **Fig. 3** in dark turquoise, which is corrected by assuming a linear drift within the pulses (light turquoise). A square wave is fitted to the data (orange) to determine the step height that ultimately corresponds to the cantilever deflection. The same procedure is performed for the nominally set 75, 149, 224, 298, and 373 mT field amplitudes, measured with ±5 mT precision. Knowing the deflection value ($\Delta z$), one can extract the resulting mechanical torque $T$ exerted by a single SAF NPs at a given magnetic field strength using equation (1), where the SAF NP is located at $x = a$, and $\Delta z = z_{B \neq 0} - z_{B=0}$.

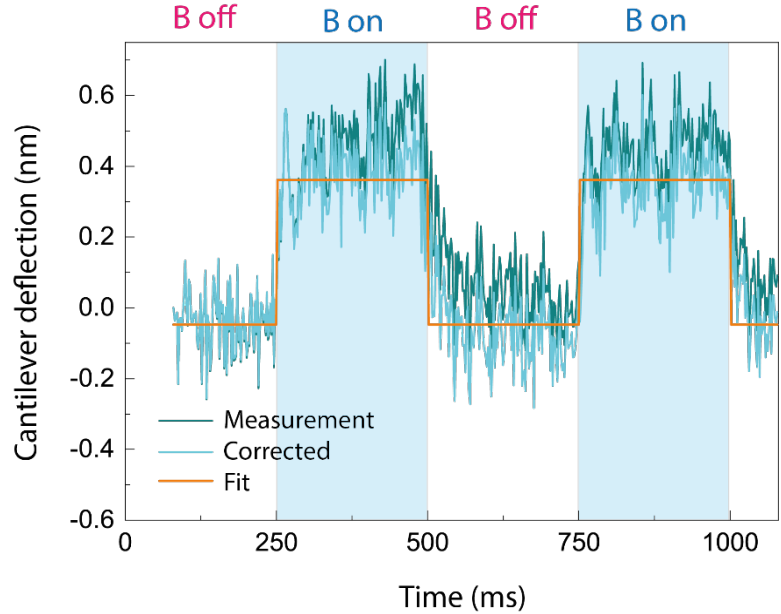

**FIG. 3**. Time dependence of the difference between the deflections of the AFM cantilever with the single SAF NP and a nominally identical empty AFM cantilever in an external magnetic field, both datasets are the average of 300 measurements (dark turquoise) with 373 mT amplitude applied in two 250 ms pulses, in between which the field is turned off. Deflection correction assuming a linear drift within the pulses is performed (light turquoise), and a square wave is fitted to the data (orange).

To be able to compare the experimentally obtained $T$ value with the theoretically predicted one, the theoretical maximum torque must be calculated using the following equation:

$$\vec{T} = \vec{m}_{\text{eff}} \times \vec{B} \qquad (2),$$

where $\vec{m}_{\text{eff}}$ and $\vec{B}$ are the effective magnetic moment of the SAF NP and the applied magnetic field, respectively.

To describe the $\vec{m}_{\text{eff}}$ for a given $\vec{B}$, we use the Stoner-Wohlfarth (SW) model[15], where the total energy of the system is determined based on a macrospin model. The two $Co_{60}Fe_{20}B_{20}$ layers, with similar saturation magnetization $M_s$, in a basic stack unit repeated five times are treated as two single spins. In order to model the magnetization state of the SAF NPs, the energy of the system is minimized (see the **supplementary material** for more details). In the SW model, $\vec{m}_{\text{eff}}$ can be determined by vectorially adding the magnetic moments $\vec{m}_i$ of the FM layers:

$\vec{m}_{\text{eff}} = \vec{m}_1 + \vec{m}_2$ (3).

**Fig. 4(a)** depicts $\vec{m}_{\text{eff}}$ for the anti-parallel/AP (left) and parallel/P (middle and right) configuration of the two coupled layers with the individual magnetic moments $\vec{m}_1$ and $\vec{m}_2$. For the AP configuration, $|\vec{m}_{\text{eff}}|$ is much smaller than for the P configuration, while the angle between $\vec{m}_{\text{eff}}$ and $\vec{B}$ is much larger. Note that the mechanical torque is generated by the deviation of the $\vec{m}_{\text{eff}}$ from the easy axis as indicated by the angle $\theta$ due to the PMA of the Co/Pt interfaces. To model the $\vec{m}_{\text{eff}}$ dependence on the applied magnetic field $\vec{B}$, we take angle $\alpha = 12\pm2°$, as it is the case for our AFM setup (**Fig. 1(b)**). **Fig. 4(b)** shows the corresponding hysteresis loop. Note that we plot $M_\alpha$, which is the projection of $\vec{m}_{\text{eff}}$ along the magnetic field direction. Thus, when $M_\alpha/M_s = 0$, there is no magnetization along the magnetic field direction, and when $M_\alpha/M_s = 1$, the magnetization is saturated and lies in the magnetic field direction. The magnetic moment has a small component along the magnetic field before switching, and after switching, it does not immediately reach saturation. Brown's paradox[34] is not easily implemented in the SW model, and hence, the switches at ±100 mT and ±200 mT are added manually, to emulate to what is measured experimentally.

Knowing the dependence of $\vec{m}_{\text{eff}}$ on the external magnetic field $\vec{B}$, the mechanical torque $\vec{T}$ generated by the SAF NP can be derived (see the **supplementary material** for more details):

$\vec{T} = m_{\text{eff}}(\vec{B}, \alpha) \cdot B_k \cdot \sin\theta \cdot \cos\theta$ (4),

where $B_k$ is the anisotropy field following from the equation for the effective anisotropy constant:

$B_k = \frac{2K}{M_s}$ (5).

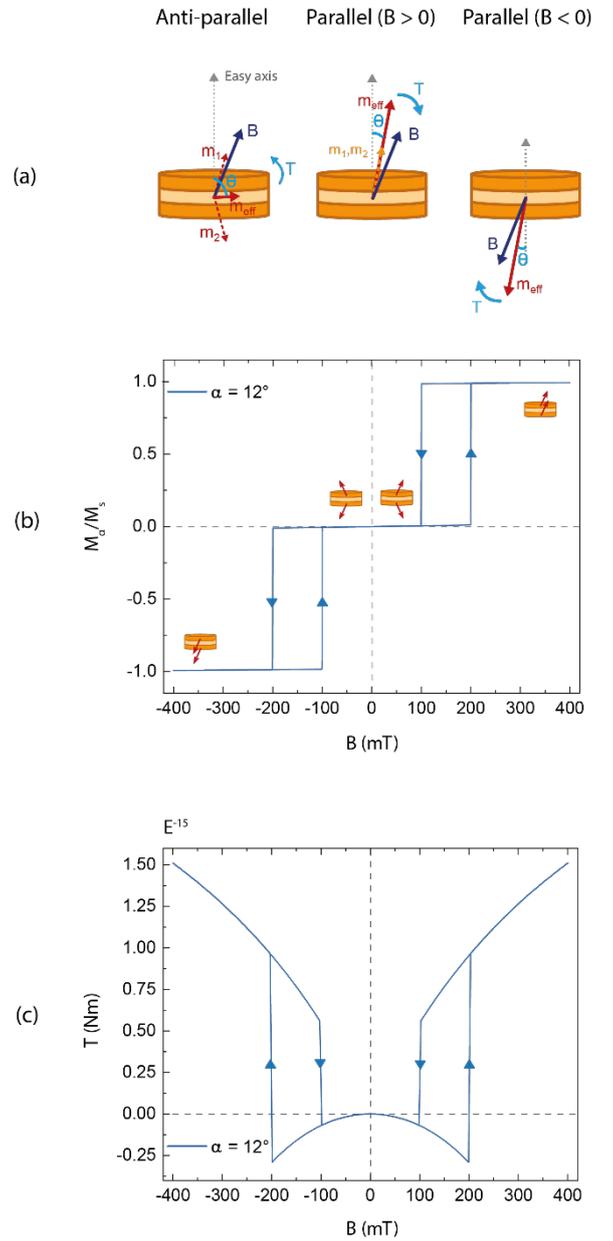

**FIG. 4**. Illustration of the Stoner-Wohlfarth (SW) model used for predicting the magnetic torque generated by SAF NPs and describing their effective magnetic moment $\vec{m}_{\text{eff}}$. (a) Schematic representation of the $\vec{m}_1$, $\vec{m}_2$, $\vec{m}_{\text{eff}}$ and resulting torque $\vec{T}$ for anti-parallel and parallel configurations of the two magnetic layers under the applied magnetic field $\vec{B}$, $\theta$ is the angle between the effective magnetic moment $\vec{m}_{\text{eff}}$ of the SAF NP and the easy axis of the platelet; (b) Hysteresis loop for a magnetic field applied under $\alpha = 12°$ with respect to the easy axis according to the SW model, red arrows indicate magnetic moments of individual ferromagnetic layers ($\vec{m}_1$, $\vec{m}_2$); (c) magnetic field dependence of the magnetic torque for a magnetic field applied under $\alpha = 12°$ with respect to the easy axis according to the SW model (equation 4). Arrows in (b) and (c) depict the field sweep direction. The switches at ±100 and ±200 mT are manually added to the model.

For our SAF NPs, $M_s$ = 1400±50 kA/m and $K$ = (389±16) kJ/m³, leading to $B_k$ = 0.56 T. These values are similar to the values found in [31] (see the **supplementary material** for more details).

In **Fig. 4(c)**, we plot the expected magneto-mechanical torque using the SW model as a function of magnetic field, where a different sign of the torque for the AP and P cases can be seen. The symmetry around $x = 0$ can be understood by looking at **Fig. 4(a)**. As there is no preference for $\vec{m}_1$ and $\vec{m}_2$ to point either up or down along the easy axis, changing the sign of the $\vec{B}$ effectively results in a 180° rotation of both $\vec{B}$ and $\vec{m}_{\text{eff}}$, which does not influence the direction of the $\vec{T}$, and thus the sign remains the same. The derivatives in these areas are also different. Before switching, $\vec{m}_{\text{eff}}$ increases as $\vec{m}_1$ and $\vec{m}_2$ slightly cant towards the magnetic field, resulting in a slope around zero magnetic field. After switching, however, $\vec{m}_{\text{eff}}$ is already near saturation, and the canting decreases the angle between the $\vec{B}$ and $\vec{m}_{\text{eff}}$. To limit this effect and generate the maximum $\vec{T}$, a high PMA is required.

From equation (4), the torque is maximized for $\theta$ equal to 45° leading to:

$$T_{max} = \frac{mB_k}{2} \quad (6)$$

Since $m = M_s \cdot V$ = 3.11·10⁻¹⁴ Am², $T_{max}$ = 8.6·10⁻¹⁵ Nm. In our experiment, the angle α between the cantilever and the horizontal plane (the same as the angle between the easy axis and the external magnetic field $\vec{B}$) is equal to 12±2°, which influences $m_{\text{eff}}$ as a function of $\vec{B}$ and $\alpha$. Since we only have a few nm deflection with our AFM cantilever, we can neglect the change in angle ($\alpha$) as a function of field. At a 12-degree angle, the maximum torque in this configuration is therefore expected to be ≈ 1.5·10⁻¹⁵ Nm at 373±5 mT field (**Fig. 4(c)**).

**Fig. 5** presents the main findings of our work where the experimentally measured torques calculated according to equation (1) and theoretically predicted torques according to the SW model (equation (4), orange line) are plotted together showing excellent agreement. Note that the measurements were performed for two nominally identical AFM MLCT-B cantilevers, each containing one SAF NP close to the tip (green and cyan dots), which are in good agreement with each other. The difference between the two we attribute to slight (±2°) variations in the angle the NPs make with the magnetic field after drop casting. Furthermore, the SW model lies within the 95% confidence interval of both measurements, confirming that the magnetic torque generated by SAF NPs is efficiently translated into a mechanical torque. To determine the resulting maximum force, we can assume the arm of the torque to be the radius of the SAF NP (9.4·10⁻⁷ m), and for the maximum obtained torque of 1.6± 0.6·10⁻¹⁵ Nm (the average of the two measurements and a 95% confidence interval), we find $F = T/r \approx$ 1.7·10⁻⁹ N (1700 pN) at the SAF NP apex. This measured torque and derived force of the SAF NPs is strong enough for most applications in magnetomechanical cell actuation, e.g., can be used to rupture (cancer) cell membranes and lysosomes, change protein-lipids interactions, contribute to unfolding protein molecules, etc. Moreover, design of the SAF NPs opens a route for easy tuning of the built-in magnetic anisotropy and size, reducing the torque and allowing for much smaller mechanical stimuli for e.g. ion channel activation, which only requires 1-10 pN[35–38].

In conclusion, we have developed a method to measure the mechanical torques from individual SAF NPs based on the AFM, which is a widely available characterization technique. The measured torques were found to be in line with the SW model, and large enough for the destruction of cancer cells. Given that the AFM sensitivity could be further improved, e.g., by adding modulation techniques or operating in dynamical tip modes, this work opens a straightforward and widely applicable method for characterizing the torques generated by nm-sized SAF NPs and other anisotropic magnetic nanostructures.

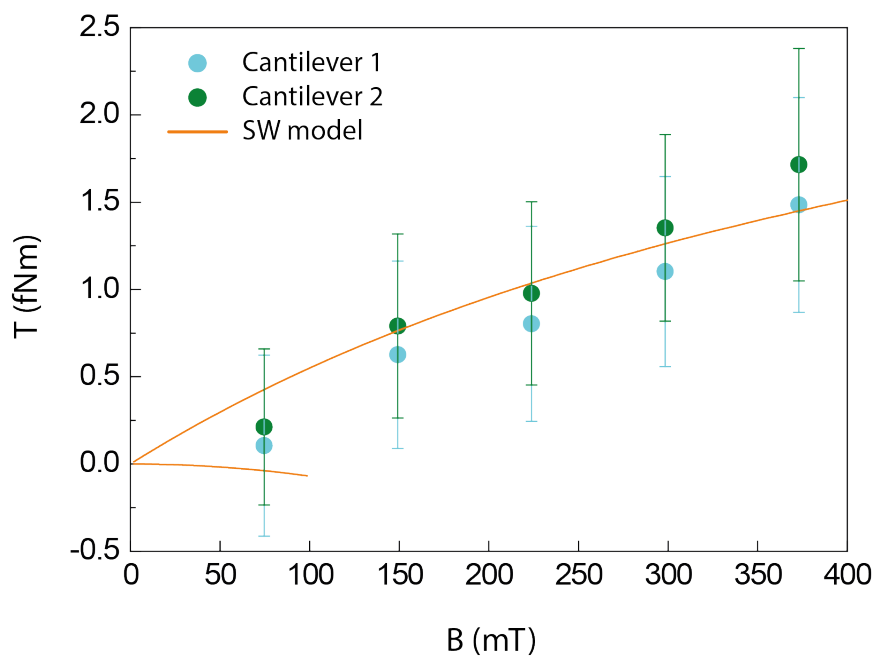

**FIG.5**. Results of the torque magnetometry of two nominally identical MLCT-B AFM cantilevers, each containing one SAF NP close to the tip (green and cyan dots), where the torques are calculated according to equation (1) versus theoretically predicted torques according to the Stoner-Wohlfarth (SW) model (orange line). The error bars represent the 95% confidence intervals for both measurements.

See the supplementary material for details of the AFM setup and cantilevers specifications, the derivation of the point force and point torque equations, details of the torque magnetometry (cantilever calibration, deflection measurement, torque calculation), magnetic properties of SAF NPs, and derivations of the SW model equations.

The authors gratefully acknowledge Mark de Jong and Jeroen Francke from Eindhoven University of Technology, and Sergey Lemeshko from Bruker Nano Surfaces for their kind assistance with the AFM setup, measurement protocols, and the cantilever choice.

We thank the financial support from the ICMS CRT-STA-CSSF-RL Synthetic project. M.V.E. gratefully acknowledges the funding from the European Union's Horizon 2020 research and innovation programme under the Marie Skłodowska-Curie grant agreement no. 899987.

**AUTHOR DECLARATIONS**
**Conflict of Interest**
The authors have no conflicts to disclose.

**Author Contributions**

Maria V. Efremova: Conceptualization (equal); Funding acquisition (equal); Data curation (equal); Formal analysis (equal); Investigation (supporting); Methodology (equal); Validation (equal); Writing – original draft (equal); Writing – review & editing (equal). Lotte Boer: Conceptualization (equal); Data curation (equal); Formal analysis (equal); Investigation (equal); Methodology (equal); Validation (equal); Writing – original draft (equal); Writing – review & editing (equal). Laurenz Edelmann: Data curation (supporting); Formal analysis (supporting); Methodology (supporting); Validation (supporting); Writing – review & editing (equal). Lieke Ruijs: Conceptualization (supporting); Formal analysis (supporting); Methodology (supporting); Validation (supporting); Writing – review & editing (equal). Jianing Li: Conceptualization (supporting); Formal analysis (supporting); Methodology (supporting); Validation (supporting); Writing – review & editing (equal). Marc A. Verschuuren: Methodology (supporting); Validation (supporting); Writing – review & editing (equal). Reinoud Lavrijsen: Conceptualization (equal); Funding acquisition (equal); Investigation (equal); Methodology (equal); Project administration (equal); Supervision (equal); Validation (equal); Writing – review & editing (equal).

**DATA AVAILABILITY**

The data that support the findings of this study are available within the article and its supplementary material.

# Supplementary Material:
# AFM Cantilever Magnetometry for Measuring Femto-Nm Torques Generated by Single Magnetic Particles for Cell Actuation

**Maria V. Efremova[1*], Lotte Boer[1*], Laurenz Edelmann[1], Lieke Ruijs[1], Jianing Li[1], Marc A. Verschuuren[2], and Reinoud Lavrijsen[1]**

* equal contribution

[1]Department of Applied Physics, Eindhoven University of Technology, P.O. Box 513, 5600 MB Eindhoven, The Netherlands

[2]SCIL Nanoimprint Solutions, High Tech Campus 11, 5656EA Eindhoven, the Netherlands

## 1. AFM Setup and cantilevers specifications

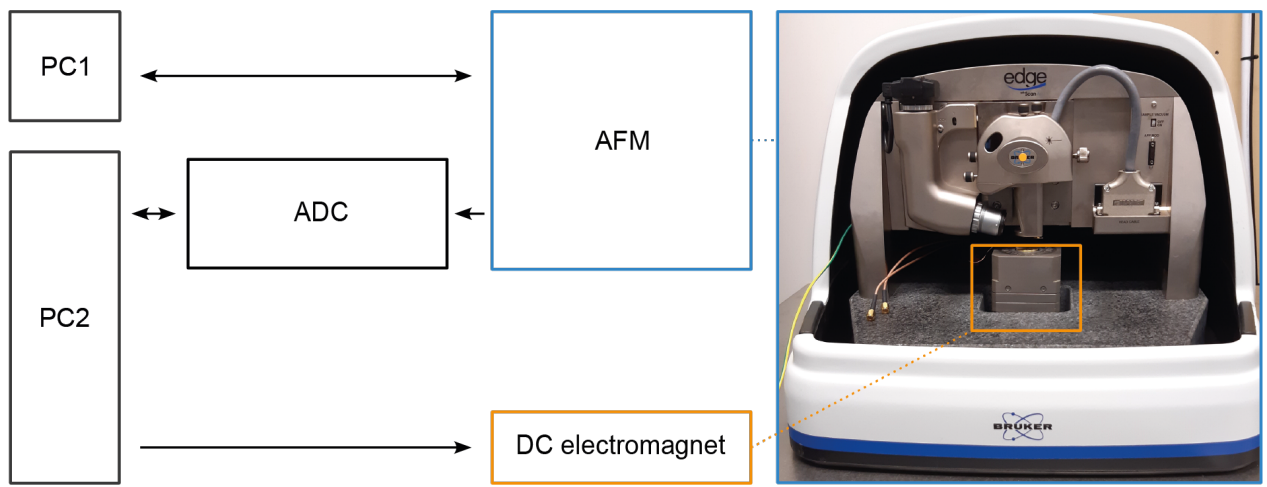

**Fig. S1.** Schematic diagram of the AFM setup used for the cantilever magnetometry. The central component of the setup is the Bruker Dimension Edge AFM (blue frame), which is fully controlled via the NanoDrive software on the first computer (PC1, black frame). The cantilever is mounted on the AFM scanhead via a non-magnetic cantilever holder (Bruker DCHNM). GMW 5203 vertical projection field magnet is installed underneath the cantilever (DC electromagnet, orange). The magnet is controlled via LabView software with a custom-written script on the second computer (PC2, black frame). PC2 also controls the analog-to-digital converter (ADC) connected to the AFM, which was used to synchronize the timing of the AFM tip deflection signal and the applied magnetic field pulses.

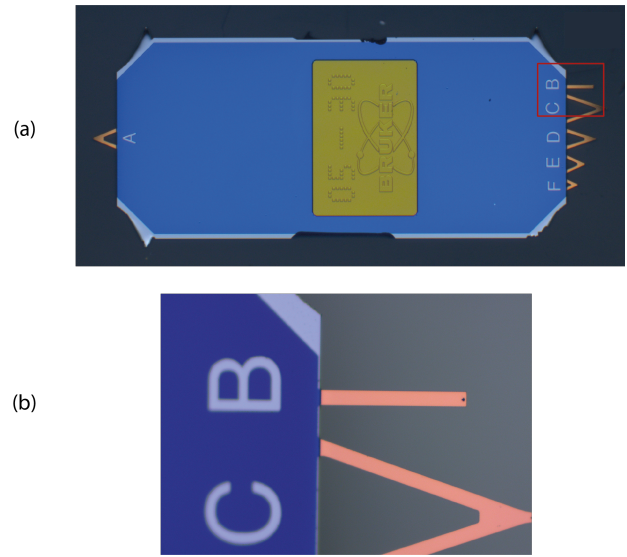

**Fig. S2.** Optical microscopy image of an AFM MLCT cantilever chip. (a) Overview of the full chip with multiple cantilevers; (b) zoom-in of the rectangular MLCT-B cantilever used in the current study, also shown by the red frame in a).

**Table S1**. Nominal parameters of an AFM cantilever MLCT-B used in the current study

| Parameter | Value |
| --- | --- |
| Spring constant ($k$) | 0.02 N/m |
| Length ($L$) | 210 µm |
| Tip location ($x_{tip}$) | 205 µm |
| Width ($\omega$) | 20 µm |
| Thickness ($t$) | 0.55 µm |

**2. Derivation of the point force at the tip and point torque at the arbitrary position of the rectangular AFM cantilever**

In both cases (point force and point torque), the bending of the cantilever at a distance x from the clamped end (cf. **Fig. 1(b)**) is related to the torque at the same position $T(x)$ via the geometrical moment of inertia ($I$ [kg·m²]) and Young's modulus ($E$ [GPa]) of the cantilever as derived in Adhikari *et al.*, 2012:[1]

$$T(x) = \frac{EI}{R} \tag{S1}$$

where the product $EI$ represents the rigidity of the cantilever, and $R$ is the cantilever bending across its length. In this equation, we assume a cantilever where the length is much larger than the width ($L \gg \omega$), such that the bending across its width can be neglected. Since our chosen cantilever satisfies these requirements (**Supplementary Table S1**), we only consider deflections, $\Delta z$, which are much smaller than $L$. $1/R$ can then be approximated as $\frac{d^2z}{dx^2}$, and the reduced arm of the torque due to cantilever bending can be neglected.[2] To verify this assumption, Hooke's law can be used to approximate the deflection as $\Delta z = F/k$.

Filling in the typical $k$ = 0.02 N/m for our cantilevers and the maximum force of ~2 nN estimated in the main text, we find a deflection of 100 nm, which is indeed several orders of magnitude smaller than the 210 μm corresponding to the cantilever length. Combining the two assumptions, Equation (S1) can then be written as:

$$T(x) = EI \frac{d^2z}{dx^2} \tag{S2},$$

which proves that the cantilever bending increases with increased torque and decreases for more rigid or stiffer cantilevers.

## 2.1 Point force at the tip of the rectangular AFM cantilever

Determining the deflection due to a point force at the tip 1) is applicable to measuring force-distance curves for calibration of the AFM measurement, and 2) it allows us to write the rigidity ($EI$) of our cantilever in terms of known parameters. The deflection due to a point force at the tip is shown schematically in **Fig. S3**.

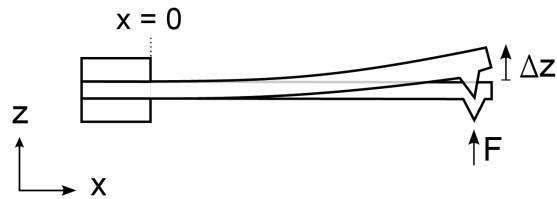

**Fig. S3**. Schematic depiction of the bending of a rectangular AFM cantilever due to a point force $F$ applied at the tip; $\Delta z$ is the cantilever deflection in the coordinate system $x - z$; $x = 0$ marks the clamped end of the cantilever, while $z = 0$ is the top of the cantilever when no force is applied.

When a point force $F$ is applied at $x = a$, the position-dependent torque corresponds to $T(x) = (x - a)F$ for $0 \leq x \leq a$ and $T(x) = 0$ for $x > a$. For $0 \leq x \leq a$, Equation (S2) becomes:

$$\frac{d^2z}{dx^2} = \frac{F(x-a)}{EI} \tag{S3}.$$

Integrating this twice over $x$ with boundary conditions $\frac{dz}{dx} = 0$ and $z = 0$ at $x = 0$ gives the position-dependent deflection of the cantilever:

$$z = \frac{F}{6EI}(x^3 - 3x^2 a) \tag{S4}.$$

This equation for the cantilever deflection requires knowledge of its rigidity, which is the product of its geometrical moment of inertia ($I$) and Young's modulus ($E$). Whereas the geometrical moment of inertia is known for rectangular cantilevers, their Young's modulus is not well known and ranges from 56 to 370 GPa in literature for Silicon Nitride films with a thickness in the same order of magnitude as the used cantilever (t = 0.55 μm).[3] However, the rigidity of the cantilever can be related to known parameters by looking at the deflection of the tip for a point force at the same location ($x = a = x_{tip}$), which is given by:

$$z_{tip} = z(x = x_{tip}) = -\frac{F}{3EI} x_{tip}^3 \tag{S5}$$

Approximating the cantilever as an ideal spring with $F = kz$, and substituting z with Equation S5 results in:

$$z_{tip} = -\frac{F}{3EI} x_{tip}^3 = -\frac{F}{k} \tag{S6}$$

From this, the rigidity of the cantilever follows to be:

$$EI = \frac{k x_{tip}^3}{3} \tag{S7}$$

which can be used in the previously derived equations for the cantilever deflection. The geometrical moment of inertia for rectangular cantilevers is given by:[4]

$$I = \frac{1}{12} \omega t^3 \tag{S8}$$

with $\omega$ and $t$ its width and thickness, respectively. In combination with Equation (S7), the Young's modulus is then found to be:

$$E = \frac{4k x_{tip}^3}{\omega t^3} \tag{S9}$$

For verification, we can fill in the nominal values of our cantilevers as stated in **Supplementary Table S1**, and find the Young's modulus to be 160 GPa. This falls within the range of 56-370 GPa described in the literature for films with thicknesses similar to that of out cantilever,[3] and thus we conclude that we can use Equation (S9).

**2.2 Point torque at the arbitrary position of the rectangular AFM cantilever**

For the torque generated by a single synthetic antiferromagnetic nanoplatelet (SAF NP), we assume the platelet to be rigidly bound to the cantilever via electrostatic interactions. Moreover, the platelet diameter, 1.88 µm, is much smaller than the length of the cantilever, 210 µm, and thus we consider the torque of the platelet as a point torque acting on the cantilever. This is depicted schematically in **Fig. S4**.

The torque $T$ of the platelet acts on the cantilever between the clamped end ($x = 0$) and its location ($x = a$). In this region, $T(x) = T$, whereas for $x > a$, $T(x) = 0$; thus, no bending occurs, and the slope of the cantilever remains constant. Therefore, for $0 \leq x \leq a$, equation (S3) becomes:

$$\frac{d^2 z}{dx^2} = \frac{T}{EI} \tag{S10}$$

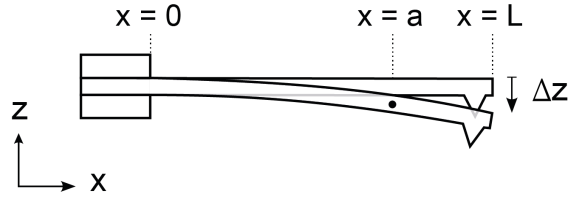

**Fig. S4.** Schematic depiction of the bending of a rectangular AFM cantilever due to a point torque $T$ applied at $x = a$; $\Delta z$ is the cantilever deflection in the coordinate system $x - z$; $x = 0$ marks the clamped end of the cantilever, while $z = 0$ is the top of the cantilever when no torque is applied.

Using the same boundary conditions as for the point force, $\frac{dz}{dx} = 0$ and $z = 0$ at $x = 0$, this equation can be integrated twice to find:

$$z(x) = \frac{T}{2EI}x^2 \tag{S11}$$

for $0 \leq x \leq a$. Taking into account the continuity of the deflection and the slope at $x = a$, we find:

$$z(x) = z_a + \frac{dz}{dx}\bigg|_{x=a}(x-a) = \frac{T}{2EI}(2x - a^2) \tag{S12}$$

From this equation, it is clear that the slope of the deflection changes linearly with the location of the applied torque. To gain further insight in the cantilever bending, deflection profiles are plotted for various torque positions. The result is shown in **Fig. S5**, where the torque is taken to be $9 \cdot 10^{-15}$ Nm as the estimated value of a single platelet (see Equation (7) in the main text), and the cantilever parameters from **Supplementary Table S1** are used.

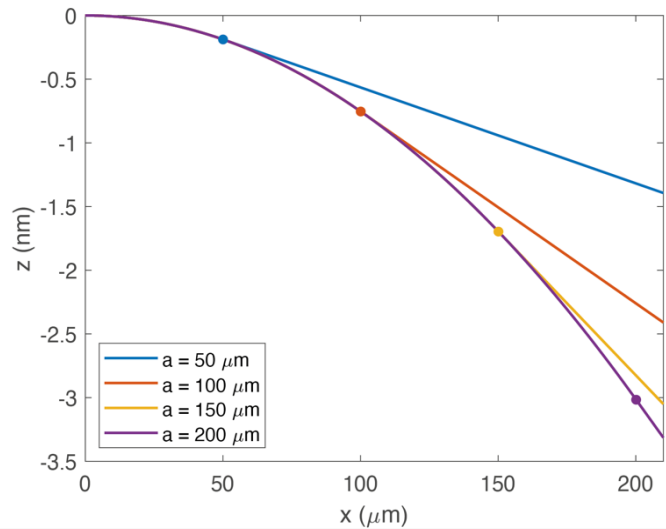

**Fig. S5.** Cantilever bending due to point torque of $9 \cdot 10^{-15}$ Nm according to equation (S12) applied at different points on the MLCT-B cantilever. Lines denote the cantilever, with dots corresponding to the point of the torque application (the total cantilever length is 210 μm, while the tip position $x_{tip}$ is 200 μm). Note the difference in scale between the y (nm) and x (μm) axis.

Note the order of magnitude of the deflection, which is significantly smaller than the cantilever length, as initially assumed. Looking at the figure, the magnitude of the deflection is significantly impacted by the location of the point force. Particularly, when the torque is applied closer to the clamped end, substantial larger deflection is observed. To achieve the maximum deflection signal for a single platelet, it must be placed towards the free end of the cantilever. Moreover, we find the maximum deflection for single platelets to be on the nm scale, which is large enough to be measured using the AFM, and small enough to ignore significant angle differences relative to the magnitude of the magnetic fields we apply, which can measure deflections down to the order of 1 Å.[5]

## 3. Torque magnetometry

### 3.1 Calibration measurements

Before performing torque magnetometry using the AFM, the cantilever's deflection sensitivity and spring constant must be calibrated. The force-distance (FD) curve determination and thermal noise spectra are integrated into the Nanodrive software of the Bruker AFM. For the FD curves, a piece of silicon wafer is used as a substrate. An example of the resulting curve is shown in **Fig. S5(a)**. The deflection sensitivity is determined by fitting a straight line through the steep part of the curve and taking the inverse of the slope. The value is found to be around 0.2 μm/V for most measurements.

As will be discussed in Section 2 above, the cantilever deflection is different for the point torque compared to the point force. As the deflection sensitivity is calibrated using a point force, we must correct this by taking into account the position of the laser on the cantilever ($x$) in Equations (S4) and (S11). We can determine the correction factor ($A$) via

$$A = \frac{\Delta z_{PT}(x_{tip})}{\Delta z_{PT}(x_{laser})} \Big/ \frac{\Delta z_{PF}(x_{tip})}{\Delta z_{PF}(x_{laser})} \qquad (S13).$$

where $\Delta z_{PT}$ and $\Delta z_{PF}$ correspond to the deflections due to the point torque and point force, respectively. Knowing the position of the laser spot, the correction factor can be determined from Equations (S4) and (S11)

To determine the spring constant, both the Sader method[6] and the thermal tune method[7] can be used. In this study, we chose the Sader method because it aims to standardize the spring constant calibration throughout different labs and gives values close to the nominal spring constant of 0.02 N/m for the MLCT-B cantilever.

To calibrate the spring constant using the Sader method, first, a thermal noise spectrum is measured on the AFM, after which the resonance peak is fitted with a Lorentzian peak, as shown in **Fig. S5(b)**. From this fit, the resonance frequency and Q-factor ($f/\Delta f$) of the cantilever are determined. The Sader Method GCI[8] provides a web page where reference data is uploaded by AFM users, and the spring constant of cantilevers is calculated based on this data and the measured Q-factor and resonance frequency of the cantilever. The spring constant is determined after depositing a SAF NP onto the cantilever, where no significant difference is detected due to the presence of the NP and/or drop casting procedure.

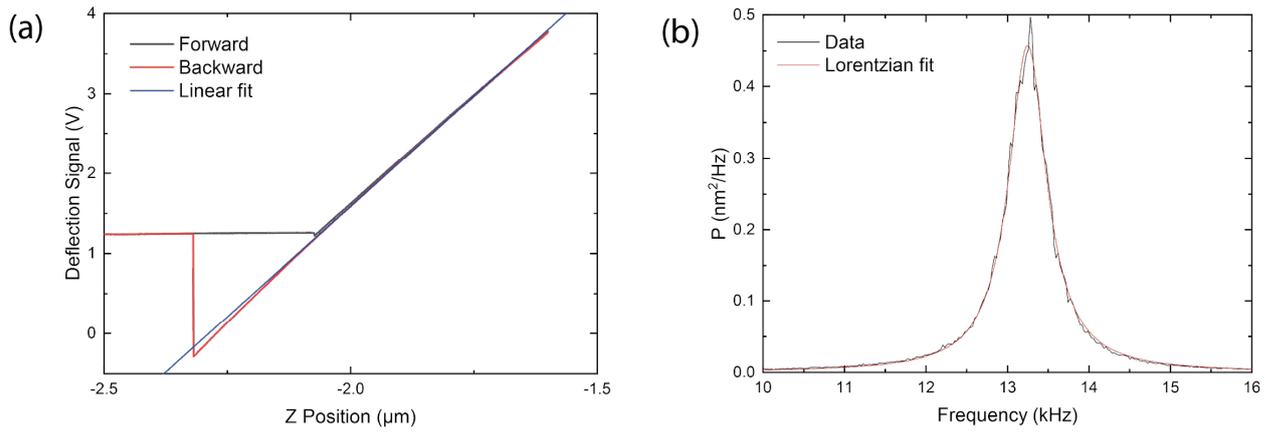

**Fig. S6**. Calibration measurements with fits for an MLCT-B cantilever. (a) Typical force-distance curve; (b) thermal tune spectrum showing the resonance frequency.

## 3.2 Cantilever deflection measurement in a magnetic field

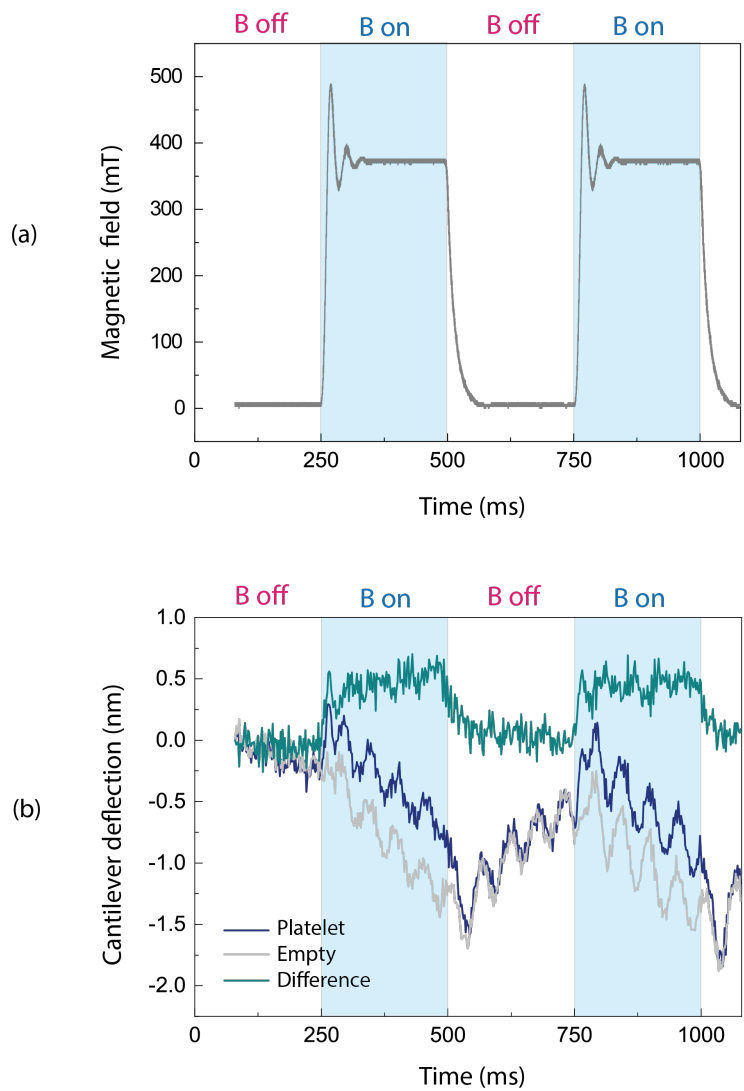

**Fig. S7**. Details of the pulsed field measurement scheme in the SAF NPs torque magnetometry. (a) Example of a single pulsed field sequence (time dependence of the magnetic field amplitude). The measurement starts with a 170 ms equilibration time, then two pulses of the same field amplitude (373 mT) are applied with a 250 ms pause in between, and 80 ms of the equilibration time is added at the end. This pulse and pause length 1) allow the field to stabilize (which in both cases takes ≈ 100 ms) and 2) are short enough to not experience major nonlinear drift effects; (b) Time-dependent deflections of the AFM cantilever with the single SAF NP (blue), a nominally identical empty AFM cantilever (grey), and their difference (dark turquoise, same curve is displayed in **Fig. 3**) in an external magnetic field, both datasets are averaged over 300 measurements with 373 mT amplitude applied in two 250 ms pulses, in between which the field is turned off. In blue and dark turquoise curves, a clear jump occurs when the field is turned on around 200 ms. Note that the signal shows an oscillation with ≈ 200 Hz frequency. This oscillation is apparent for both cantilevers while not being visible in the difference curve and, hence, is not caused by the SAF NP. As the oscillation is not averaged out, it is likely to originate from the electromagnetic coil inductance (RC circuit) that starts when the magnetic field is turned on.

**3.3 Torque calculation**

After determining the signals for the different magnetic field strengths, we can calculate the torque according to Equation (S12). As described in Section 3.1, we need to correct for the difference in bending between the point force and the single platelet torque. During the alignment, the laser was located around the center of the cantilever, thus it was chosen to use a position of 122±38 µm, from which the correction factor in Equation (S13) is 1.22±0.09 µm. As the cantilever is 210 µm long, this corresponds to a range of the laser spot between 0.35L and 0.7L. An image of the laser spot aligned on a cantilever taken with the camera in the AFM is shown in **Fig. S8**.

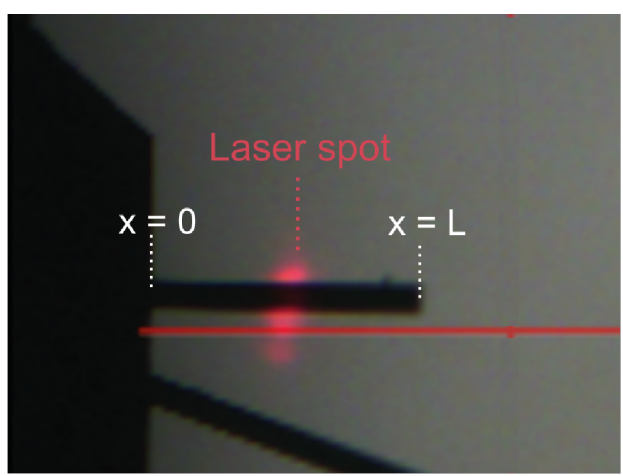

**Fig. S8.** Laser spot positioned on the AFM cantilever.

## 4. Synthetic antiferromagnetic nanoplatelets (SAF NPs)

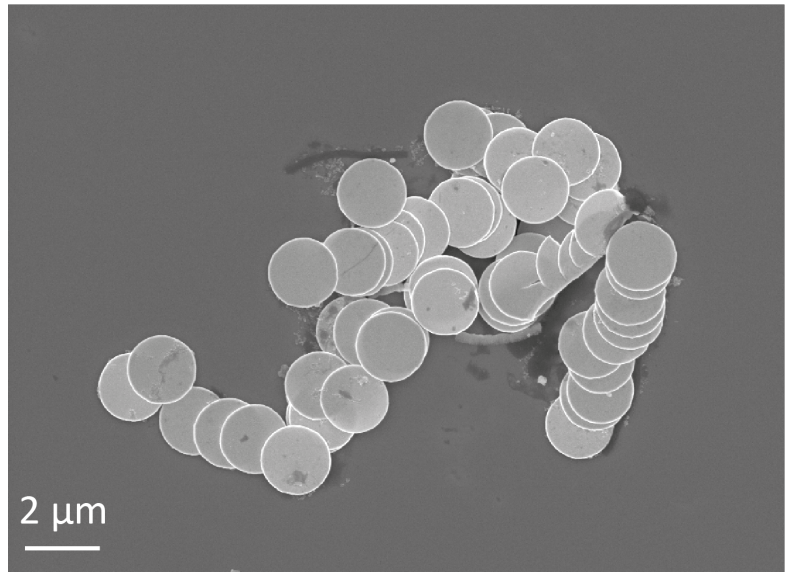

**Fig. S9.** Scanning electron microscopy image of the SAF NPs with 1.88 μm diameter and 72 nm thickness dropcasted on the AFM cantilever MLCT-B.

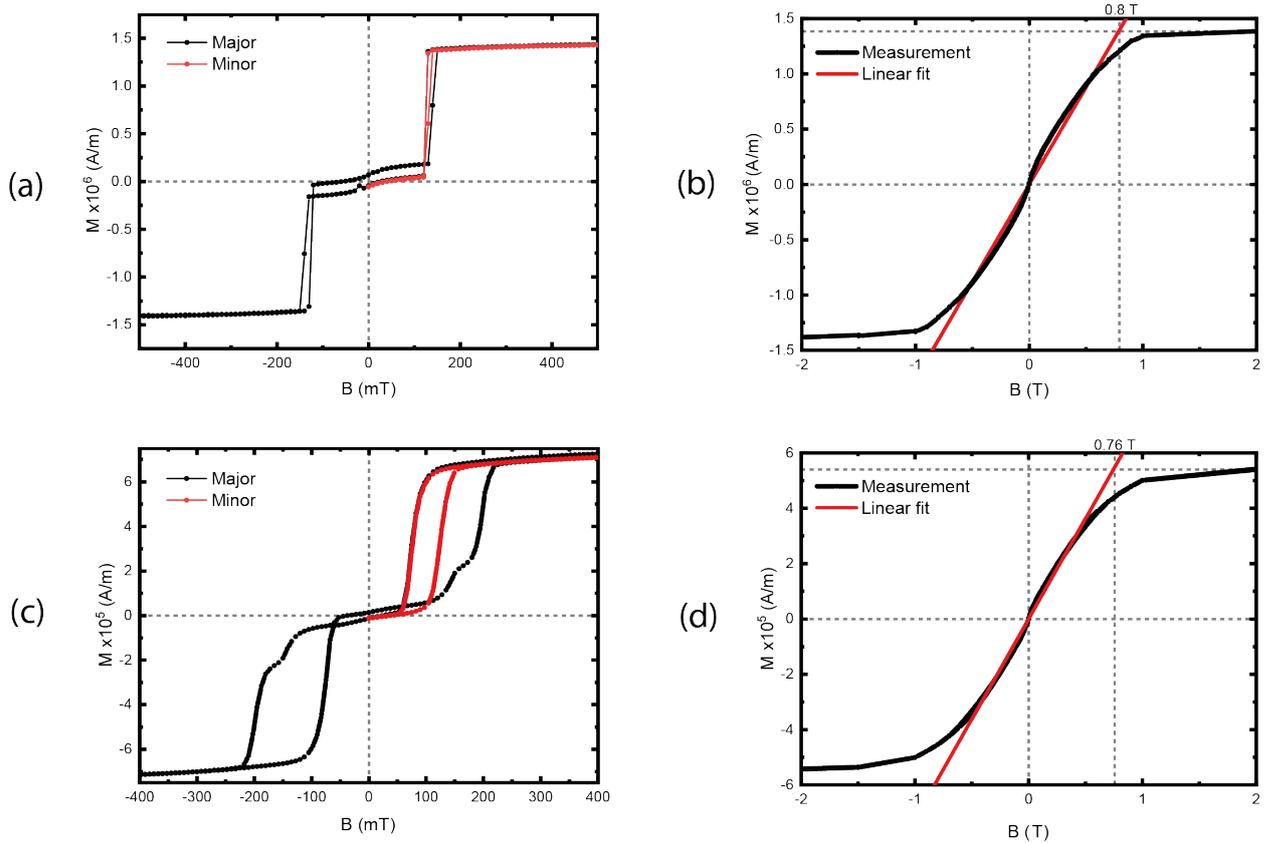

**Fig. S10**. Magnetic characterization of the SAF structures with the VSM-SQUID. (a), (b) Hysteresis loops of SAF thin films measured along the easy (a) and hard (b) axes; (c), (d) hysteresis loops of SAF NPs dropcasted on a piece of a silicon wafer measured along the easy (c) and hard (d) axes. The red lines in (b) and (d) represent the linear fit used to determine the saturation field. A linear correction was applied to compensate for the diamagnetic background.

Table S3. Parameters of the SAF NPs used to model their magnetic torque

| Parameter | Value |
|---|---|
| Radius (R) | $0.94*10^{-6}$ m |
| Thickness of a single $Co_{60}Fe_{20}B_{20}$ layer (t) | $8.00*10^{-9}$ m |
| Magnetic volume of a single $Co_{60}Fe_{20}B_{20}$ layer (V) | $2.20*10^{-21}$ m$^3$ |
| Coupling energy per unit area (J) | $2.08*10^{-4}$ J/m$^3$ |
| Effective anisotropy constant (K) | $3.90*10^{5}$ J/m$^3$ |
| Saturation magnetization ($M_s$) | $1.40*10^{6}$ A/m |

## 5. Stoner -Wohlfarth model

### 5.1 Energy considerations

Stoner-Wohlfarth (SW) model[9] related to the total energy of the system is used in this work to describe the effective magnetic moment $\vec{m}_{eff}$ of the ferromagnetic layers in SAF NPs for a given magnetic field $\vec{B}$. SW model is a macrospin model, treating the two ferromagnetic (FM) layers as two single spins.

In order to model the magnetization state of the platelets, we minimize the energy of the system, from which the $\vec{m}_{eff}$ can be predicted. The angles used in the SW model are shown schematically in **Fig. S11(a) (left)**. As mentioned before, the first contribution to the energy comes from the magnetic anisotropy, and is given by

$$E_a = KV\sin^2(\theta) \tag{S14},$$

where $K$ corresponds to the effective anisotropy constant [J/m$^3$], $V$ is the magnetic volume of the layer, and $\theta$ is the angle of the magnetic moment with respect to the easy axis of the SAF NP. This energy term is minimum when $\theta = 0, \pi$, which corresponds to the magnetic moment aligning with the easy axis of the SAF NP. The second contribution to the energy is the Zeeman energy, given by

$$E_z = -\vec{m} \cdot \vec{B} = -mB\cos(\alpha - \theta) \tag{S15},$$

where $\alpha$ is the angle between the magnetic field and the easy axis of the SAF NP. This contribution becomes minimum when the magnetic moment aligns with the applied magnetic field, competing with the anisotropy when the magnetic field is applied under an angle. The last energy term to take into account is the RKKY energy:

$$E_{RKKY} = J_{RKKY}V/t \cos(\theta_2 - \theta_1) \tag{S16},$$

where $J_{RKKY}$ is the coupling energy per unit area in [J/m$^2$] and $\theta_i$ is the angle of the magnetic moment in the respective layer with respect to the easy axis, and $t$ is the thickness of a magnetic layer. For positive $J_{RKKY}$, the RKKY energy is minimum for $\theta_2 - \theta_1 = \pi$, thus favoring anti-parallel alignment of the magnetic moments.

(a)
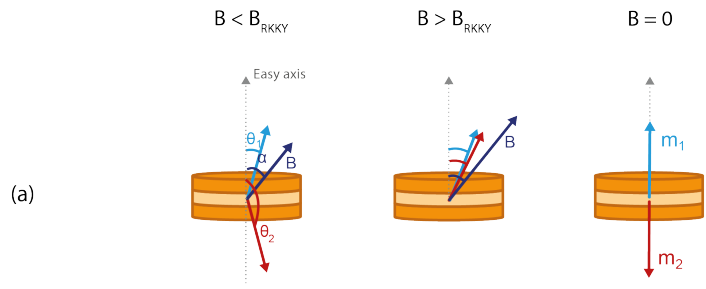

(b)
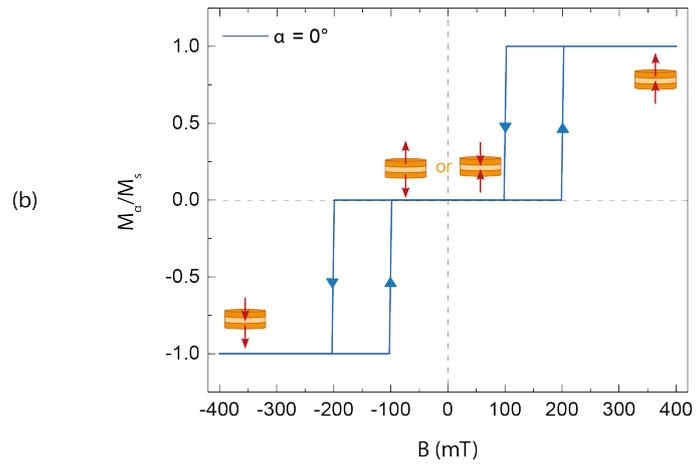

(c)
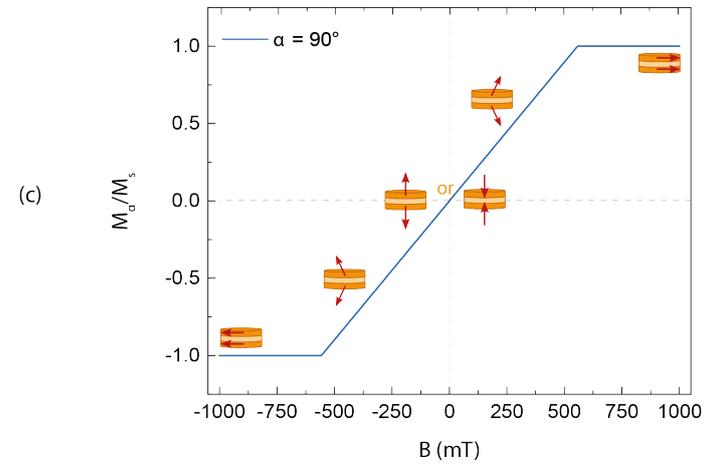

(d)
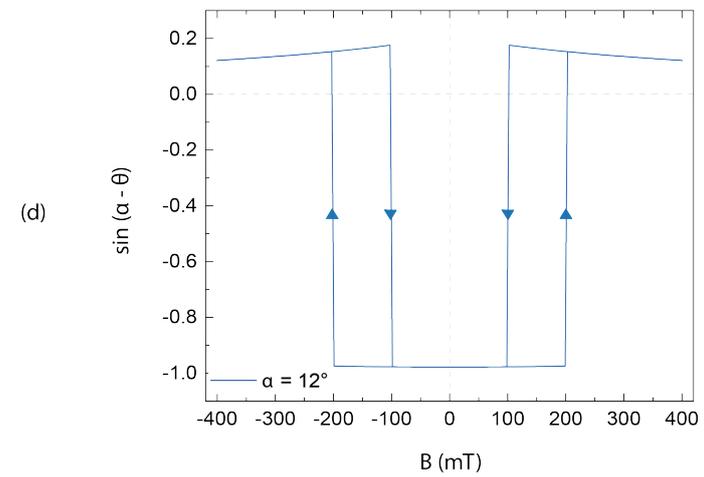

**Fig. S11**. Illustration of the Stoner-Wohlfarth (SW) model used for predicting the magnetic torque generated by SAF NPs and describing their effective magnetic moment $\vec{m}_{\text{eff}}$. (a) Schematic representation of the magnetic moments in the $Co_{60}Fe_{20}B_{20}$ layers for different magnetic field magnitudes. The light blue and red arrows represent the magnetic moments of the two layers, with their corresponding angles $\theta_1$ and $\theta_2$, $B_{RKKY}$ is the coupling field of one FM layer on the other; (b), (c) Hysteresis loops for a magnetic field applied (b) along the easy axis ($\alpha$ = 0) and (c) along the hard axis ($\alpha = \pi/2$) of the SAF stack according to the SW model for typical values of the SAF NPs (cf. **Supplementary table S3**), red arrows indicate magnetic moments of individual ferrimagnetic layers ($\vec{m}_1$, $\vec{m}_2$); (d) magnetic field dependence of the $\sin(\alpha - \theta)$ term in Equation (S21) for a magnetic field applied under $\alpha$ = 12° with respect to the easy axis according to the SW model. Arrows in (b) and (c) depict the field sweep direction. The switches at ±100 and ±200 mT are manually added to the model.

To describe the total energy of the SAF NPs, the anisotropy and Zeeman energy has to be taken into account for both FM layers, together with the RKKY coupling between them. Note that the two $Co_{60}Fe_{20}B_{20}$ magnetic layers in a SAF NP are equal in thickness, possess the same anisotropy constant and possess equal magnetic moments, i.e., $t_1 = t_2 = t$, $K_1 = K_2 = K$, and $m_1 = m_2 = m$. This allows us to write the total energy per unit volume as:[10]

$$\frac{E}{V} = K[\sin^2(\theta_1) + \sin^2(\theta_2)] + \frac{J}{t}\cos(\theta_2 - \theta_1) - M_S B[\cos(\alpha - \theta_1) + \cos(\alpha - \theta_2)] \quad (S17),$$

where $M_S$ is the saturation magnetization. Note that due to the Brown's paradox,[9] one cannot model the switching behavior, and the switches are manually added to the SW model to mimic the switching behavior as physically measured in real samples at room temperature and quasi-static field sweeping conditions.

By minimizing Equation (S17), the angles $\theta_1$ and $\theta_2$ at equilibrium can be found, and the effective magnetic moment, $\vec{m}_{\text{eff}}$, can be determined by vectorially adding the magnetic moments $\vec{m}_i$ of the FM layers (**Fig. S11(b)**):

$$\vec{m}_{\text{eff}} = \vec{m}_1 + \vec{m}_2 \quad (S18).$$

The hysteresis loops resulting from the SW model for a magnetic field applied along the easy axis ($\alpha$ = 0) and the hard axis ($\alpha = \pi/2$) are shown in **Figs. S11(b)** and **S11(c)**. There, we plot $M_\alpha$, which is the projection of $\vec{m}_{\text{eff}}$ on the magnetic field axis. Thus, when $M_\alpha/M_s$ = 0, there is no magnetization along the magnetic field direction, and when $M_\alpha/M_s$ = 1, the magnetization is saturated and lies in the magnetic field direction.

In this study, we investigate the SAF stack where $K >> J/t$. Therefore, when a magnetic field smaller than the coupling field on one FM layer on the other ($B_{RKKY}$) is applied along the easy axis (**Fig. S11(b)**), the magnetic moments remain anti-parallel, resulting in no effective magnetic moment. However, when the absolute value of the magnetic field becomes large enough, the RKKY coupling can be overcome, and both magnetic moments align with the magnetic field, resulting in saturation. When the magnetic field is decreased again, the magnetic moments switch back to anti-parallel for a smaller magnetic field than initially, giving the typical hysteresis behavior of a ferromagnet. When a magnetic field is applied along the hard axis (**Fig. S11(c)**), the magnetic moments are pulled in the direction of the magnetic field due to the Zeeman interaction. When the magnetic field is large enough to overcome the anisotropy energy and RKKY coupling, the magnetic moments align with the magnetic field, and saturation magnetization is reached.

If we consider the angle α = 12±2°, as is the case for our AFM setup (**Fig. 1(a)**), the SAF NPs behavior becomes a combination of the easy and hard axis cases, as depicted schematically in **Fig. 4(b)**. When no magnetic field is applied (**Fig. 4(b), right**), the magnetic moments align in an anti-parallel fashion along the easy axis of the platelet. When the magnitude of the field is increased (**Fig. 4(b), left**), the magnetic moments cant towards the magnetic field and remain largely anti-parallel, resulting in only a small net moment along the magnetic field. Only when the magnetic field is large enough to overcome the RKKY coupling (**Fig. 4(b), middle**), the angles $\theta_1$ and $\theta_2$ become similar, resulting in a relatively large effective magnetic moment compared to the anti-parallel configuration.

## 5.2 Torque derivation

Magnetic torque of the SAF NPs is given by

$$\vec{T} = \vec{m}_{\text{eff}} \times \vec{B} \tag{S19}.$$

Thus, the magnetic torque is only present when there is an angle between $\vec{B}$ and $\vec{m}_{\text{eff}}$. Due to the perpendicular magnetic anisotropy (PMA), the magnetic moment wants to stay out-of-plane which results in a rotation of the SAF NP where the magnetic torque is translated to a mechanical torque. Using the energy density following from the SW model, an expression for the mechanical torque can be derived. As the anisotropy is strong with respect to the RKKY-coupling (K >> J/t), the angles $\theta_1$ and $\theta_2$ are similar. For that reason, the average angle of the two magnetic moments, $\theta$, is used, as is illustrated in **Fig. S11(b)**. Here, we make an assumption that every $Co_{60}Fe_{20}B_{20}$ layer has the same magnetic anisotropy (which in reality, may be different due to the sample fabrication process). Hence, the energy density given in Equation (S17) can be simplified for a system with just the net magnetic moment:

$$\frac{E}{V} = -M_S B \cos(\alpha - \theta) + K\sin^2\theta \tag{S20}.$$

For this situation, the magnitude of the magnetic torque following from Equation (S19) is equal to:

$$\vec{T} = m_{eff}(\vec{B}, \alpha) * B\sin(\alpha - \theta) \tag{S21}.$$

By setting the derivative with respect to $\theta$ of Equation (S20) to zero an expression for $sin(\alpha - \theta)$ can be found:

$$\sin(\alpha - \theta) = \frac{2K}{M_S B}\sin\theta\cos\theta = \frac{B_k}{B}\sin\theta\cos\theta \tag{S22}.$$

Knowing the term $\sin(\alpha - \theta)$, the torque generated by a SAF NP can be determined based on the SW model and combining equations (S17) and (21). **Fig. S11(d)** shows the resulting

$\sin(\alpha - \theta)$ term in equation (S21) when magnetic field $\vec{B}$ is applied under α = 12° with respect to the easy axis.